\documentclass[conference]{IEEEtran}
\IEEEoverridecommandlockouts

\usepackage{cite}
\usepackage{amsmath,amssymb,amsfonts}
\usepackage{algorithmic}
\usepackage{graphicx}
\usepackage{textcomp}
\usepackage{xcolor}
\usepackage{acro}
\usepackage{url}
\usepackage{siunitx}
\usepackage{booktabs}
\usepackage{tabularx}
\usepackage{multirow}
\usepackage{todonotes}
\usepackage{subfigure}

\usepackage{hyperref}
\hypersetup{
    colorlinks = true,
    allcolors=blue,
    urlcolor=black,
    linkcolor=black
}

\usepackage{xspace}
\let\OldTexttrademark\texttrademark
\renewcommand{\texttrademark}{\OldTexttrademark\xspace}%

\DeclareSIUnit{\tbps}{Tb/s}
\DeclareSIUnit{\tb}{Tb}
\DeclareSIUnit{\gbps}{Gb/s}
\DeclareSIUnit{\gb}{Gb}
\DeclareSIUnit{\mpps}{Mpps}

\newcolumntype{Y}{>{\centering\arraybackslash}X}

\def\BibTeX{{\rm B\kern-.05em{\sc i\kern-.025em b}\kern-.08em
    T\kern-.1667em\lower.7ex\hbox{E}\kern-.125emX}}

\usetikzlibrary{svg.path}
\definecolor{orcidlogocol}{HTML}{A6CE39}
\tikzset{
    orcidlogo/.pic={
            \fill[orcidlogocol] svg{M256,128c0,70.7-57.3,128-128,128C57.3,256,0,198.7,0,128C0,57.3,57.3,0,128,0C198.7,0,256,57.3,256,128z};
            \fill[white] svg{M86.3,186.2H70.9V79.1h15.4v48.4V186.2z}
            svg{M108.9,79.1h41.6c39.6,0,57,28.3,57,53.6c0,27.5-21.5,53.6-56.8,53.6h-41.8V79.1z M124.3,172.4h24.5c34.9,0,42.9-26.5,42.9-39.7c0-21.5-13.7-39.7-43.7-39.7h-23.7V172.4z}
            svg{M88.7,56.8c0,5.5-4.5,10.1-10.1,10.1c-5.6,0-10.1-4.6-10.1-10.1c0-5.6,4.5-10.1,10.1-10.1C84.2,46.7,88.7,51.3,88.7,56.8z};
        }
}
\newcommand\orcidicon[1]{\href{https://orcid.org/#1}{%
        \mbox{\begin{tikzpicture}[yscale=-1,transform shape, scale=0.03] % Change scale here
                \pic{orcidlogo};
            \end{tikzpicture}}}}

\usepackage{fancyhdr}
\fancypagestyle{IEEEtitlepagestyle}{
    \fancyhf{} % clear all header and footer fields
     
    \fancyhead[C]{
        \small This work has been accepted at the \emph{4th KuVS Workshop on Network Softwarization (KuVS NetSoft)} under the Creative Commons Attribution 4.0 International License (CC BY 4.0).
        \textit{Digital Object Identifier \href{http://dx.doi.org/10.15496/publikation-105106}{10.15496/publikation-105106}}
        }
}

\newcommand\citeN\cite

\newcommand\sect[1]{Section~\ref{sec:#1}}

\newcommand\tabl[1]{Table~\ref{tab:#1}}

\newboolean{makevspace}
\newcommand{\cvspace}[1]{%
   \ifthenelse
   {\boolean{makevspace}}
   {\vspace{#1}}
   {}%
}
\acsetup{single}

\DeclareAcronym{P4}{
    short = P4,
    long = Programming Protocol-independent Packet Processors,
    single = P4
}

\DeclareAcronym{P4TG}{
    short = P4TG,
    long = P4-based traffic generator,
}

\DeclareAcronym{TG}{
    short = TG,
    long = traffic generator,
}

\DeclareAcronym{RTT}{
    short = RTT,
    long = round trip time,
}

\DeclareAcronym{TX}{
    short = TX,
    long = TX,
}

\DeclareAcronym{RX}{
    short = RX,
   long = TX,
}

\DeclareAcronym{L1}{
    short = L1,
    long = Layer 1,
}

\DeclareAcronym{L2}{
    short = L2,
    long = Layer 2,
}

\DeclareAcronym{L3}{
    short = L3,
    long = Layer 3,
}

\DeclareAcronym{IAT}{
    short = IAT,
    long = inter-arrival time,
}

\DeclareAcronym{MPLS}{
    short = MPLS,
    long = Multiprotocol Label Switching,
    first-style = short,
}

\DeclareAcronym{ARP}{
    short = ARP,
    long = Address Resolution Protocol,
    first-style = short,
}

\DeclareAcronym{MAT}{
    short = MAT,
    long = match-action table,
}

\DeclareAcronym{CBR}{
    short = CBR,
    long = constant bit-rate,
}

\DeclareAcronym{DUT}{
    short = DUT,
    long = device under test,
}

\DeclareAcronym{SR}{
    short = SR,
    long = segment routing,
    first-style = short,
}

\DeclareAcronym{SRv6}{
    short = SRv6,
    long = segment routing v6,
}

\DeclareAcronym{LSE}{
    short = LSE,
    long = label stack entry,
    long-plural-form = label stack entries,
}

\DeclareAcronym{IMIX}{
    short = IMIX,
    long = Internet mix,
}

\DeclareAcronym{NDP}{
    short = NDP,
    long = neighbor discovery protocol,
    first-style = short,
}

\DeclareAcronym{NETCONF}{
    short = NETCONF,
    long = Network Configuration Protocol,
}

\DeclareAcronym{UI}{
    short = UI,
    long = user interface,
    first-style = short,
}

\DeclareAcronym{VxLAN}{
    short = VxLAN,
    long = Virtual extensible LAN,
    first-style = short,
}

\DeclareAcronym{PM}{
    short = PM,
    long = precision mode,
}

\DeclareAcronym{RM}{
    short = RM,
    long = rate mode,
}

\DeclareAcronym{ILP}{
    short = ILP,
    long = integer linear program,
}

\DeclareAcronym{Mpps}{
    short = Mpps,
    long = mega-packets per second,
}

\begin{document}

\title{Enhancements to P4TG: Protocols, Performance, and Automation}

\author{
    \IEEEauthorblockN{Fabian Ihle$^{\orcidicon{0009-0005-3917-2402}}$, Etienne Zink$^{\orcidicon{0009-0001-0879-535X}}$, Steffen Lindner$^{\orcidicon{0000-0002-5274-4621}}$, Michael Menth$^{\orcidicon{0000-0002-3216-1015}}$}
    \IEEEauthorblockA{University of T\"ubingen, Chair of Communication Networks
        \\Email: \{fabian.ihle, etienne.zink, steffen.lindner, menth\}@uni-tuebingen.de}
    \thanks{The authors acknowledge the funding by the Deutsche Forschungsgemeinschaft (DFG) under grant ME2727/3-1. This work was supported by the bwNET 2.0 project which is funded by the Ministry of Science, Research and the Arts Baden-Württemberg (MWK). The authors alone are responsible for the content of this paper.}
}

\maketitle

\begin{abstract}
    P4TG is a hardware-based \ac{TG} running on the Intel Tofino\texttrademark 1 ASIC and was programmed using the programming language P4.
    In its initial version, P4TG could generate up to $\mathbf{10 \times} \qty[text-series-to-math]{100}{\gbps}$ of traffic and directly measure rates, packet loss, and other metrics in the data plane.
    Many researchers and industrial partners requested new features to be incorporated into P4TG since its publication in 2023.
    With the recently added features, P4TG supports the generation of packets encapsulated with a customizable VLAN, QinQ, VxLAN, MPLS, and SRv6 header.
    Further, generation of IPv6 traffic is added and P4TG is ported to the Intel Tofino\texttrademark 2 platform enabling a generation capability of up to $\mathbf{10 \times} \qty[text-series-to-math]{400}{\gbps}$.
    The improvement in user experience focuses on ease of operation.
    Features like automated \acs{ARP} replies, improved visualization, report generation, and automated testing based on the IMIX distribution and RFC~2544 are added.
    %Combining those enhancements prepares P4TG for the ever increasing demands of future networks.
    Future work on P4TG includes \acs{NDP} to facilitate IPv6 traffic, and a \acs{NETCONF} integration to further ease the configuration.
\end{abstract}


\vspace{-0.3cm}

\section{Introduction}
\label{sec:introduction}

\Acfp{TG} are used to test network devices or applications.
Typically, \acp{TG} generate frames with different sizes, rates, traffic patterns, and packet headers to mirror the behavior of real-world traffic.
For network testing, \acp{TG} further implement measurements of \acs{TX}/\acs{RX} rates, \acp{RTT}, packet loss, and out-of-order packets.
\Acp{TG} can either be implemented in software or hardware.
They come with several different trade-offs.
Software-based \acp{TG} are cost-effective, customizable, and run on commodity hardware~\cite{trex}.
However, they are not as accurate and performant as hardware-accelerated \acp{TG} that use specialized hardware components~\cite{EmGa17, CoVo24}.
These components result in better performance and higher accuracy while being less flexible and more expensive.

Programmable switches facilitate the development of hardware-supported tools at affordable cost which was the initial motivation for the development of the \acf{P4TG}~\cite{LiHae23}.
%Therefore, the \ac{P4TG} was developed.
\Ac{P4TG} leverages the capabilities of the Intel Tofino\texttrademark switching ASIC for traffic generation in Ethernet/IP networks.

In its initially published version, \ac{P4TG} supported customizable Ethernet and IPv4 headers.
However, to effectively test productive networks, more diverse traffic patterns are needed.
They include enhanced traffic generation capabilities such as new protocols and higher data rates.
In this paper, we give an overview of the enhancements to \ac{P4TG}, including VLAN, QinQ, VxLAN, SRv6 and MPLS encapsulation, IPv6 support, and up to \qty{4}{\tbps} of traffic generation.
Further, the enhancements to \ac{P4TG} improve the user experience by facilitating automated network testing through report generation and test profiles.
This demonstrates another advantage of a self-programmed hardware-based \ac{TG}, namely ease of extensibility.
\section{The Traffic Generator P4TG}
The traffic generator \ac{P4TG} is a hardware-based \acs{TG} implemented for the Intel Tofino\texttrademark\ switching ASIC~\cite{LiHae23}.
The source code of \ac{P4TG} is open-source on GitHub~\cite{p4tg-git}.
\Ac{P4TG} features custom traffic generation and measurement in the data plane.
A generated packet in \ac{P4TG} contains a UDP header, a \ac{P4TG} header, and a customizable Ethernet and IPv4 header. 
The \ac{P4TG} header contains a sequence number, transmission timestamp, and stream ID for measurements.
\Ac{P4TG} measures rates, frame types, sizes, as well as lost and out-of-order packets in the data plane.
Further, \acp{RTT} and \acp{IAT} are sampled in the control plane.
Various frame sizes ranging from \qty{64}{\byte} (byte) to \qty{9000}{\byte} are supported.
The frames are padded with random data if the frame size exceeds the configured cumulative header size.

For traffic generation, the internal traffic generation ports of the Intel Tofino\texttrademark are leveraged~\cite{PublicTNA}.
Up to two additional loopback ports per physical port are required for traffic customization and measurement.
Therefore, \ac{P4TG} can use ten physical ports for traffic generation on a 32-port switch enabling a bandwidth of \qty{1}{\tbps}.
The generated traffic is either \ac{CBR} or Poisson (random) traffic.
Up to seven \acs{CBR} streams or one Poisson stream are configurable for traffic generation.
\section{Enhancements to P4TG}
\label{sec:novelties}

Since its publication in 2023, \ac{P4TG} has been extended with many features requested by academic and industrial partners.
This section provides an overview of the new features, explains them in detail, and divides them into two areas: enhanced traffic generation capabilities and improved user experience.
%%%%%%%%%%%%%%%%%%%%%%%%%%%%%%%%%%%%%%%%%%%%%%%%%%%%%%%%%%%%%%%%%%%%%%%%%%%%%%%%%%%%%%%%%%%
%%%%%%%%%%%%%%%%%%%%%%%%%%%%%%%%%%%%%%%%%%%%%%%%%%%%%%%%%%%%%%%%%%%%%%%%%%%%%%%%%%%%%%%%%%%

\subsection{Overview}
\label{sec:overview}

\tabl{overview} provides a non-exhaustive overview of existing, new, experimental, and planned features of \ac{P4TG}.
The following sections introduce the new and experimental features in more detail.
New features are implemented and merged into the stable branch of \ac{P4TG}.
Experimental features are implemented but need more testing until they are merged.
Planned features are future work that will be included later on.

\begin{table}[t]
    \caption{\ac{P4TG} feature overview.}
    \label{tab:overview}
    \centering
    \resizebox{\columnwidth}{!}{%
        \begin{tabularx}{1.2\columnwidth}{l ll ll}
            \toprule
            \textbf{Existing} & \multicolumn{2}{c}{\textbf{New}}  & \textbf{Experimental} & \textbf{Planned} \\\midrule
            • Ethernet        & • IPv6         & • VLAN          & • Test profiles       & • \acs*{NDP}     \\
            • IPv4            & • QinQ         & • \acs*{VxLAN}  & • File reporting      & • \acs*{NETCONF} \\
            • Statistics      & • \acs*{MPLS}  & • \acs*{SRv6}   & • Auto. testing       &                  \\
            • Tofino 1        & • \acs*{ARP} replies & • Tofino 2  & • Localization       &                  \\
            • Web frontend    & • Rust backend & • Dark mode     &                       &                  \\
            • Python backend  &               &                 &                       &                  \\
            \bottomrule
        \end{tabularx}%
    }
    \vspace{-0.4cm}    
\end{table}

Adding features to P4TG is a threefold process.
First, new packet formats require changes to packet parsing and processing in the P4 data plane.
Second, the controller must be extended including configuration of the new data plane functionality and modification of the REST API.
Third, the new features must be made available in the frontend by adding forms to configure them using the REST API.

With Intel's decision to open-source the Intel Tofino SDE, P4TG can now be compiled and used by everyone who has an Intel Tofino\texttrademark\ ASIC~\cite{p4open}.
We provide a GitHub Continuous Integration/Continuous Deployment (CI/CD) workflow that automatically compiles P4TG for the Intel Tofino\texttrademark\ 1/2 on updates to the P4TG GitHub repository.

%%%%%%%%%%%%%%%%%%%%%%%%%%%%%%%%%%%%%%%%%%%%%%%%%%%%%%%%%%%%%%%%%%%%%%%%%%%%%%%%%%%%%%%%%%%
%%%%%%%%%%%%%%%%%%%%%%%%%%%%%%%%%%%%%%%%%%%%%%%%%%%%%%%%%%%%%%%%%%%%%%%%%%%%%%%%%%%%%%%%%%%

\subsection{Enhanced Traffic Generation Capabilities}
\label{sec:technicalFeatures}

This section describes the traffic generation capabilities added to \ac{P4TG}.

%%%%%%%%%%%%%%%%%%%%%%%%%%%%%%%%%%%%%%%%%%%%%%%%%%%%%%%%%%%%%%%%%%%%%%%%%%%%%%%%%%%%%%%%%%%

\subsubsection{Traffic Customization}
\label{sec:encapsulations}
In the initial published version, \ac{P4TG} supported the generation of Ethernet and IPv4 traffic with source and destination address randomization.
The randomization allows the generation of multiple IP streams emulating real network traffic.
\Ac{P4TG} is extended with \textit{IPv6} traffic generation capabilities.
For IPv6, up to 48 bits can be randomized in the source and destination addresses.

Various traffic encapsulation protocols are necessary to model the network conditions in real-world environments.
\textit{VLAN} is an encapsulation that adds a VLAN tag to an Ethernet frame.
The use cases for VLAN tags are manifold including segmentation of physical networks into multiple logical domains, traffic engineering, QoS, and traffic prioritization.
Additionally, with \textit{QinQ} two VLAN headers are added to an Ethernet frame allowing for more sophisticated traffic engineering using \ac{L2} tunneling, and a larger VLAN ID space.
Further, \textit{\Ac{VxLAN}} encapsulates Ethernet frames within a UDP datagram to provide a \ac{L2} overlay in a \acs{L3} network.
\textit{\Ac{MPLS}} is a broadly established protocol in wide-area networking that enables high-performance forwarding.
%In \ac{MPLS}, an \ac{MPLS} stack, consisting of \acp{LSE}, is pushed to a packet. % at an ingress router of an \ac{MPLS} domain.
%Packets are forwarded at MPLS transit routers based on the information in the \ac{MPLS} stack.
Stacked MPLS labels enable tunneling, or encoding explicit paths leveraging the source routing paradigm in \ac{SR}-\ac{MPLS}.
With the recent advance to segment routing (SR), technologies such as \ac{SR}-\ac{MPLS} and \acs{SRv6} have become more important.
For \ac{SRv6}, a routing extension header is added to the IPv6 base header.
This routing extension header encodes an explicit path and contains a list of segments, i.e., IPv6 addresses that must be traversed.

\Ac{P4TG} supports the encapsulation of generated traffic with VLAN and QinQ headers, up to 15 MPLS \acp{LSE}, and up to three segments in \ac{SRv6}\footnote{SRv6 is only supported on the Intel Tofino\texttrademark 2.}.
Additionally, the already encapsulated traffic can be further encapsulated by a \ac{VxLAN} header\footnote{Not in combination with SRv6 encapsulation.}.
The header fields of the encapsulation protocols are fully customizable by the control plane during runtime.
By customizing the content of MPLS \acp{LSE} or SRv6 SIDs, traffic with protocol extensions such as MPLS network actions (MNA)~\cite{ietf-mpls-mna-fwk-15} and SID compression~\cite{ietf-spring-srv6-srh-compression-23} can be generated.
The VLAN encapsulation of \ac{P4TG} was used for an evaluation in~\cite{IhLi24} and the \ac{MPLS} encapsulation in~\cite{IhMe24}.

%%%%%%%%%%%%%%%%%%%%%%%%%%%%%%%%%%%%%%%%%%%%%%%%%%%%%%%%%%%%%%%%%%%%%%%%%%%%%%%%%%%%%%%%%%%

\subsubsection{Tofino\texttrademark{}\ 2 Support}
\label{sec:tofino2Support}

\Ac{P4TG} was initially implemented for the Intel Tofino\texttrademark 1 switching ASIC.
Since the publication of \ac{P4TG}, the implementation has been extended to support the more powerful Intel Tofino\texttrademark 2 hardware.
\Ac{P4TG} running on Intel Tofino\texttrademark 2 hardware supports traffic generation of up to \qty{4}{\tbps} with \SI{400}{\gbps} per port.
Furthermore, the extended pipeline size of the Intel Tofino\texttrademark 2 compared to the Intel Tofino\texttrademark 1 enables more sophisticated traffic encapsulation, such as \ac{SRv6}.
The generation accuracy is evaluated in \sect{eval}.
\subsubsection{Improved Control Plane}
\label{sec:controlPlane}

The control plane is the middleware between the data plane of \ac{P4TG} and a web-based \ac{UI} for configuration.
Using the data plane API, the control plane of \ac{P4TG} configures the Intel Tofino\texttrademark hardware, aggregates measurement data, and exposes configuration and statistic endpoints through a REST API.
%The web-based UI leverages the REST API for ease of configuration.
Initially, the control plane was implemented in Python. % to access the data plane.
To make the control plane more robust, it is entirely rewritten in the Rust programming language.
Rust is an up-to-date, fast, and memory-efficient programming language.
During the control plane rewrite of \ac{P4TG}, the \texttt{rbfrt} library was developed~\cite{ZiFl25}.
The \texttt{rbfrt} library provides bindings for the gRPC interface of the Intel Tofino\texttrademark switching ASIC to configure the data plane using the Rust language.

%%%%%%%%%%%%%%%%%%%%%%%%%%%%%%%%%%%%%%%%%%%%%%%%%%%%%%%%%%%%%%%%%%%%%%%%%%%%%%%%%%%%%%%%%%%

\subsection{Improved User Experience}
\label{sec:userExperience}

This section describes features improving the user experience of \ac{P4TG}.

%%%%%%%%%%%%%%%%%%%%%%%%%%%%%%%%%%%%%%%%%%%%%%%%%%%%%%%%%%%%%%%%%%%%%%%%%%%%%%%%%%%%%%%%%%%

\subsubsection{Frontend}
\label{sec:frontend}

\Ac{P4TG} provides a web-based frontend written in ReactJS to facilitate the monitoring and configuration of measured and generated traffic.
The frontend is extended with a live visualization of generated and received traffic rates, packet loss and RTT measurement, and frame size distribution.
The visualized data over time can be exported using the REST API for better usability.
Additionally, the configuration of packet encapsulation as described in \sect{encapsulations} is available in the frontend.
Further, a dark mode and localization are added for better accessibility.
After measurement, a report file about the analyzed traffic can be exported in PDF or CSV format.

%%%%%%%%%%%%%%%%%%%%%%%%%%%%%%%%%%%%%%%%%%%%%%%%%%%%%%%%%%%%%%%%%%%%%%%%%%%%%%%%%%%%%%%%%%%

\subsubsection{Automated ARP Replies}
\label{sec:arpReplies}

In a typical network, the \acs{ARP} table of a \ac{DUT} is filled by \ac{ARP} requests and replies.
Traffic is only forwarded if the respective ARP table entry is available.
However, \ac{P4TG} did not implement the handling of \ac{ARP} requests, and the \ac{ARP} table of a \ac{DUT} had to be filled manually.
To facilitate this, \ac{P4TG} automatically responds with a pre-configured MAC address to ARP requests.
This feature can be configured on a per-port basis.

%%%%%%%%%%%%%%%%%%%%%%%%%%%%%%%%%%%%%%%%%%%%%%%%%%%%%%%%%%%%%%%%%%%%%%%%%%%%%%%%%%%%%%%%%%%

\subsubsection{Automated Testing}
\label{sec:automatedTests}

The \ac{P4TG} REST API provides endpoints to automate network testing by configuring, starting, and stopping the traffic generation.
When running multiple individual network tests in sequence, e.g., multiple tests with different stream descriptions, the developer must script those tests by explicitly calling the \texttt{start} and \texttt{stop} REST API endpoints for each test.
To facilitate automated network testing, the \ac{P4TG} REST API is extended to accept a configuration object that defines multiple tests.
Each test consists of a stream description and a test duration.
As a result, the developer can predefine multiple tests with different stream descriptions and test durations and send them all at once in a single REST API call to \acs{P4TG}.
The \acs{P4TG} controller automatically executes the tests and provides statistics for each test in the \texttt{statistics} REST API endpoint.

%%%%%%%%%%%%%%%%%%%%%%%%%%%%%%%%%%%%%%%%%%%%%%%%%%%%%%%%%%%%%%%%%%%%%%%%%%%%%%%%%%%%%%%%%%%

\subsubsection{Test Profiles}
\label{sec:testProfiles}
Traffic profiles such as the \ac{IMIX} and benchmarking suites such as defined in RFC~2544~\cite{rfc2544} are standards for network testing.
\ac{IMIX} is a traffic profile that contains a distribution of different frame sizes, reflecting real-world conditions. RFC~2544 provides a benchmarking suite for network testing with various metrics including throughput and latency benchmarking.

Frame size distributions such as IMIX, or tests for the RFC~2544 benchmarking suite can be applied by configuring P4TG accordingly.
%For traffic generation with \acs{P4TG}, a frame size distribution can be configured to match distributions such as IMIX.
%Further, the benchmarking suite of RFC~2544 can be applied by configuring the traffic generation for each test of the suite.
However, the IMIX distribution and the individual RFC~2544 tests must be configured manually by the developer.
Therefore, \ac{P4TG} is extended with a set of predefined test profiles.
Currently, the test profiles contain an IMIX and an RFC~2544 profile.
The IMIX profile comprises a predefined stream description based on the IMIX frame size distribution.
The RFC~2544 profile supports throughput and latency benchmarking, as well as frame loss rate and reset time measurements with different frame sizes.
A developer can select a predefined test profile in the UI to test a \acs{DUT}, e.g., to automatically apply the RFC~2544 benchmarking suite.
Additional profiles may be added in the future.

\subsection{Future Work}
\label{sec:futureWork}
In its current state, traffic generation is configured and statistics are retrieved using the HTTP REST API of P4TG.
However, network operators typically configure their devices using the state-of-the-art NETCONF protocol for network management.
The NETCONF protocol leverages the YANG data modeling language to manage devices with a well-defined data model.
In the future, we want to implement a \ac{NETCONF} API using YANG models to complement the REST API for ease of network configuration.
This will allow operators to integrate \ac{P4TG} seamlessly into their existing network environment.

Automated ARP replies facilitate traffic generation in IPv4 networks.
However, for IPv6 traffic, \ac{NDP} is required which complements ARP handling.
In future work, automated \ac{NDP} handling will be added to P4TG.

\section{Generation Accuracy of P4TG}
\label{sec:eval}

We evaluate the traffic generation accuracy of P4TG.
We generate the maximum possible target rate per port, i.e., \qty{100}{\gbps} for Tofino\texttrademark\ 1 and \qty{400}{\gbps} for Tofino\texttrademark\ 2. 
We loop the traffic back to an ingress port to measure the achieved L1 rate.
\tabl{tofino_comparison} compiles the results.

\begin{table}[t]
    \caption{Maximum traffic generation rate of P4TG per port on Tofino\texttrademark\ 1 and 2 for different frame sizes.}
    \label{tab:tofino_comparison}
    \centering
        \begin{tabularx}{0.62\columnwidth}{lll}
            \toprule
            \textbf{Frame size} & \textbf{Tofino 1} & \textbf{Tofino 2} \\\midrule
            \qty{64}{\byte}   & \qty{99.37}{\gbps}  & \qty{294.00}{\gbps} \\
            \qty{128}{\byte}  & \qty{98.56}{\gbps}  & \qty{388.50}{\gbps} \\
            \qty{256}{\byte}  & \qty{99.77}{\gbps}  & \qty{396.73}{\gbps} \\
            \qty{512}{\byte}  & \qty{99.55}{\gbps}  & \qty{399.83}{\gbps} \\
            \qty{1024}{\byte} & \qty{99.41}{\gbps}  & \qty{399.72}{\gbps} \\
            \qty{1518}{\byte} & \qty{99.62}{\gbps}  & \qty{398.47}{\gbps} \\
            \bottomrule
        \end{tabularx}
        \vspace{-0.4cm}
\end{table}

P4TG achieves almost line rate for \qty{100}{\gbps} on Tofino\texttrademark\ 1 for any frame size.
However, for a target rate of \qty{400}{\gbps} on Tofino\texttrademark\ 2, the desired traffic rate is approximatly obtained only for frame sizes of \qty{256}{\byte} or larger.
When large encapsulation headers like SRv6 (\qty{48}{\byte} plus \qty{16}{\byte} per segment) or VxLAN (\qty{50}{\byte}) are utilized, frames tend to be larger, which diminishes the relevance of the reported problem.

\bibliography{bibliography/literature}

\begin{thebibliography}{10}

\bibitem{trex}
{TRex Team}.
\newblock {TRex -- Realistic Traffic Generator}.
\newblock \url{https://trex-tgn.cisco.com/}, visited on 2025-01-17.

\bibitem{EmGa17}
P.~Emmerich et~al.
\newblock {Mind the Gap - A Comparison of Software Packet Generators}.
\newblock In {\em ANCS}, pp. 191--203, 2017.

\bibitem{CoVo24}
F.~G. Costa et~al.
\newblock {PIPO-TG: Parameterizable High-Performance Traffic Generation}.
\newblock In {\em NOMS}, pp. 1--9, 2024.

\bibitem{LiHae23}
S.~Lindner et~al.
\newblock {P4TG: 1 Tb/s Traffic Generation for Ethernet/IP Networks}.
\newblock {\em {IEEE Access}}, 11:17525--17535, February 2023.

\bibitem{p4tg-git}
S.~Lindner et~al.
\newblock {GitHub: P4TG}.
\newblock visited on 2025-02-12.

\bibitem{PublicTNA}
{Intel}.
\newblock {Intel Tofino Native Architecture—Public Version}.
\newblock \url{https://github.com/barefootnetworks/Open-Tofino}.

\bibitem{p4open}
{P4.org}.
\newblock {Intel’s Tofino P4 Software is Now Open Source}.
\newblock \url{https://p4.org/intels-tofino-p4-software-is-now-open-source/ /},
  visited on 2025-01-27.

\bibitem{ietf-mpls-mna-fwk-15}
L.~Andersson et~al.
\newblock {MPLS Network Actions (MNA) Framework}.
\newblock Internet-Draft mpls-mna-fwk-15, IETF, December 2024.
\newblock WiP.

\bibitem{ietf-spring-srv6-srh-compression-23}
W.~Cheng et~al.
\newblock {Compressed SRv6 Segment List Encoding (CSID)}.
\newblock Internet-Draft spring-srv6-srh-compression-23, IETF, 2025.
\newblock WiP.

\bibitem{IhLi24}
F.~Ihle et~al.
\newblock {P4-PSFP: P4-Based Per-Stream Filtering and Policing for
  Time-Sensitive Networking}.
\newblock {\em TNSM -- Special Issue on Robust and Resilient Future
  Communication Networks}, 21(5):5273 -- 5290, 2024.

\bibitem{IhMe24}
F.~Ihle et~al.
\newblock {MPLS Network Actions: Technological Overview and P4-Based
  Implementation on a High-Speed Switching ASIC}.
\newblock {\em {ArxIv Preprint}}, October 2024.

\bibitem{ZiFl25}
E.~Zink et~al.
\newblock {Rust Barefoot Runtime (RBFRT): Fast Runtime Control for the Intel
  Tofino}.
\newblock In {\em {ArxIv Preprint}}, January 2025.

\bibitem{rfc2544}
S.~Bradner et~al.
\newblock {Benchmarking Methodology for Network Interconnect Devices}.
\newblock {RFC} 2544, IETF, March 1999.

\end{thebibliography}

% Modified style from https://gitlab.com/tanwirahmad/ieee-abrv-names to abbreviate author list to et al.
\bibliographystyle{unsrt2authabbrvpp}

\end{document}